\documentclass[aps,pre,reprint]{revtex4}
\usepackage[utf8]{inputenc}
\usepackage[english]{babel}
\usepackage{amsmath}
\usepackage{amsfonts}
\usepackage{amssymb}
\usepackage{bm}
\usepackage{graphicx}
\usepackage[caption=false]{subfig}

\begin{document}
\title{Synchronization patterns in rings of time-delayed Kuramoto oscillators}
\author{Károly Dénes}
\author{Bulcsú Sándor}
\author{Zoltán Néda}

\affiliation{ Babe\c{s}-Bolyai University, Department of Physics, 1 Kogălniceanu str., 400084 Cluj-Napoca, Romania}
\begin{abstract}
Phase-locked states with a constant phase shift between the neighboring oscillators are studied in rings of identical Kuramoto oscillators with time-delayed nearest-neighbor coupling. The linear stability of these states is derived and it is found that the stability maps for the dimensionless equations show a high level of symmetry. The size of the attraction basins is numerically investigated. These sizes are changing periodically over several orders of magnitude as the parameters of the model are varied. Simple heuristic arguments are formulated to understand the changes in the attraction basin sizes and to predict the most probable states when the system is randomly initialized.  
\end{abstract}

\maketitle

\section{Introduction}
Coupled oscillator systems are widely used as model systems in studies of self-organizing phenomena and complex behavior~\cite{strogatz2004sync}. Their widespread usage in the field of complex systems can be attributed to the fact that they are able to model phase-transitions and pattern selection in various periodic phenomena. Apart from the possibility to investigate such systems analytically, they are also appropriate for simple numerical studies. These systems were among the first ones considered computationally to understand emerging synchronization~\cite{WINFREE1967} and more recently chimera states~\cite{kuramoto2002coexistence,AbramsStrogatz2004}. One of the most famous prototype system is the Kuramoto model with phase oscillators globally coupled through a sinusoidal interaction kernel~\cite{Kuramoto}. Nowadays different versions of the original model are used for targeting different type of phenomena in complex systems~\cite{DORFLER2014}. 

One of the simplest variant of the locally coupled Kuramoto model consists of rotators placed in a ring-like topology with only nearest neighbor interaction. Though this one-dimensional topology may seem trivial, the setup has proven to produce many interesting dynamical features. The problem of critical coupling for phase locked patterns in case of heterogeneous Kuramoto-type rotators has recently been studied in detail~\cite{OCHAB2010,TILLES2011,Roy2012888,HUANG2014}. Beyond the linear stability and bifurcation analysis of states, several studies focus on the size of attraction basins belonging to the emergent phase locked patterns in rings of Kuramoto oscillators~\cite{wiley_strogatz_2006,TILLES2011,denes2019}. Instead of oscillator rings, a recent paper~\cite{PIKOVSKY2019} for example considers one dimensional chains with no periodic boundary conditions and studies the effect of heterogeneity and phase shift on the robustness of synchronized states. While the above works analyze the one-dimensional rotator system from a pure fundamental dynamical systems point of view, there are application oriented studies as well. Díaz-Guilera and Arenas~\cite{DIAZ2008} investigated the possibilities of new wireless communication protocols based on emergent phase locked patterns, while Delabays et al.~\cite{DELABAYS2016} show that synchronized patterns in rings are related to loop currents, thus linking this problem to electrical engineering. It has been shown by Daniels et al.\ that in certain cases arrays of Josephson junctions can be mapped into a Kuramoto-type system with nearest neighbor coupling~\cite{DANIELS2003}. Additionally to the already discussed analytical and numerical approaches in Kuramoto systems, the phenomenon of phase locking in coupled circuits has also been studied experimentally~\cite{MATIAS1997,WOOD2001,WATANABE1996}. The plausibility of obtaining rotating waves in locally coupled mechanical systems is discussed in papers by Qin and Zhang~\cite{QIN2009}, and by Ramírez and Alvarez~\cite{RAMIREZ2015}. 

Introducing time delay in the interaction is a natural step forward in the study of coupled oscillator systems thus making the model more realistic for practical applications. Early studies on the effect of the delay suggests that frequency synchronization and order can emerge in such systems similarly to the non-delayed case~\cite{KOUDA1981,TSANG1991,NIEBUR1991,YEUNG1999}. For locally coupled systems different oscillator types have been already considered: limit-cycle oscillators near Hopf bifurcations~\cite{dodla,BONNIN2009}, Stuart-Landau and FitzHugh-Nagumo oscillators~\cite{PERLIKOWSKI2010}. When it comes to rings of delay-coupled Kuramoto oscillators, an important contribution for a sinusoidal coupling kernel and nearest neighbor coupling is the work of Earl and Strogatz~\cite{earl}. In this seminal work the stability criterion of in-phase synchrony for arbitrary coupling functions and topology is derived. For the case of anti-phase synchronizations the work of D'Huys et al.~\cite{DHUYS2008} should be mentioned. Switching between phase-locked states in delay coupled rings under the effect of noise has been also studied~\cite{DHUYS2014}.

Our work intends to continue this research by investigating the stability of other synchronization modes as well,  being also a natural continuation of our previous studies on the attraction basins and selection of phase locked patterns in non-delayed systems~\cite{denes2019,DENES2012_2}. In the followings we focus on the stability of the phase locked states studying the changes in the basin size distribution as the parameters are varied. Since time-delayed systems are formally equivalent to infinite dimensional systems \cite{wernecke2019}, one should expect also other complex attractors as well. Studies of limit cycles, chaos, and chimera-like states in Kuramoto rings are however beyond the scope of this paper.

\section{Symmetric Phase locked states of the Kuramoto ring}
Locally coupled Kuramoto rotators with time-delayed interactions are considered on a chain with periodic boundary conditions
\begin{eqnarray}
\dot{\theta_i}(t)=\omega_0+K[\sin(\theta_{i-1}(t-\tau)-\theta_i(t))+\nonumber \\+\sin(\theta_{i+1}(t-\tau)&-\theta_i(t))]\,,\qquad
\label{eq.of_motion}
\end{eqnarray}
where $\theta_i$ is the phase of the oscillator with index $i$, $i=1,2,\ldots, N$, $\omega_0$ is the natural frequency of all oscillators, $K$ is the coupling constant and $\tau$ is the length of time delay. Together with the system size $N$ there are thus four parameters of the model. The imposed periodic condition writes as $\theta_{N+1}=\theta_{1}$ 
 and $\theta_0=\theta_N$. By rescaling the time with the delay ($t\rightarrow u=t/\tau$) we get a dimensionless system of equations with only three parameters
\begin{eqnarray}
\frac{\mathrm{d}\theta_{i}(u)}{\mathrm{d}u}=\omega+\kappa[\sin(\theta_{i-1}(u-1)-\theta_i(u))+\nonumber\\+  \sin(\theta_{i+1}(u-1)-&\theta_i(u))]\,,\qquad
\label{eq.rescaled}
\end{eqnarray}
 $\omega=\omega_0\cdot\tau$ is the dimensionless intrinsic frequency and $\kappa=K\cdot\tau$ is the dimensionless coupling.
Phase-locked pattern emerging as the solution of the above system may be written as
\begin{equation}
\label{eq.phase_locked}
\theta_i(u)=\Omega u+\varphi_i\,,
\end{equation}
$\Omega$ being a final dimensionless frequency, constant over the system and time. The $\varphi_i$ phase is needed to include all kind of synchronized patterns not only in-phase synchrony \cite{dodla,wiley_strogatz_2006,TILLES2011,Roy2012888,denes2019}. We focus on symmetric states where the $\varphi_i-\varphi_{i-1}=\Delta\phi$ terms are constant over the system.  Solution of this form of  Eq.~(\ref{eq.rescaled}) yields for the final frequency \cite{earl} the equation
\begin{equation}
\label{eq.frequency}
\Omega=\omega-2\kappa\cos(\Delta\phi)\sin(\Omega)\,,
\end{equation}
where the phase shift between neighboring oscillators, $\Delta\phi\in[-\pi,\pi)$, is equivalent to a directed distance measure on the perimeter of the unit circle~\cite{denes2019}. Due to the imposed periodic boundary conditions,  in each time-moment their sum over all $\{i,i+1\}$ oscillator pairs should be a multiple of $2\pi$. Using this sum, one can define a winding number $m$, which is considered to be an integer between $-N/2$ and $N/2-1$ for even, respectively between $-\frac{N-1}{2}$ and $\frac{N-1}{2}$ for odd number of oscillators $N$. In-phase synchrony corresponds to $m=0$, while anti-phase synchrony gets the $m=-N/2$ value (for even oscillator number). Each phase-locked state can be thus unambiguously defined by the combination of a winding number $m$ or a phase shift $\Delta\phi$ with a corresponding $\Omega$ frequency, calculated from Eq.~(\ref{eq.frequency}). 

\section{Stability of the phase-locked states}

Now we turn to the stability of the considered phase-locked states. Our approach is based on the standard linear stability method used by Earl and Strogatz to study the stability of in-phase synchrony in similar systems~\cite{earl}. Here, in contrast to previous works we refer to synchrony in a broader sense allowing states with nonzero winding numbers $m$ to appear. Using the same workaround, we consider a perturbation to the phase-locked states
\begin{equation}
\label{eq.perturbation}
\theta_i(u)=\Omega u+\varphi_i+\epsilon\,\eta_i(u)\,,
\end{equation}
where $\epsilon\ll 1$. First we concentrate on two particular states, namely in-phase and anti-phase synchrony. 

The stability of the most widely studied in-phase synchronization state $m=0$ has already been investigated in Ref.~\cite{earl}. The allowed final frequency $\Omega$ is linked to the stability of the state, that is the state is stable for $\cos(\Omega)>0$. All possible candidates for stable frequencies can be calculated from Eq.~(\ref{eq.frequency}). We point out that the derivation presented in \cite{earl}  also holds for the $m=-N/2$ case, resulting in a simple sign change in the stability condition. The final frequency for the $m=-N/2$ state is stable hence for $\cos(\Omega)<0$. A similar result is also presented for the anti-phase case in Ref.~\citep{DHUYS2008}.

\begin{figure}[t]
	\includegraphics[width=9cm]{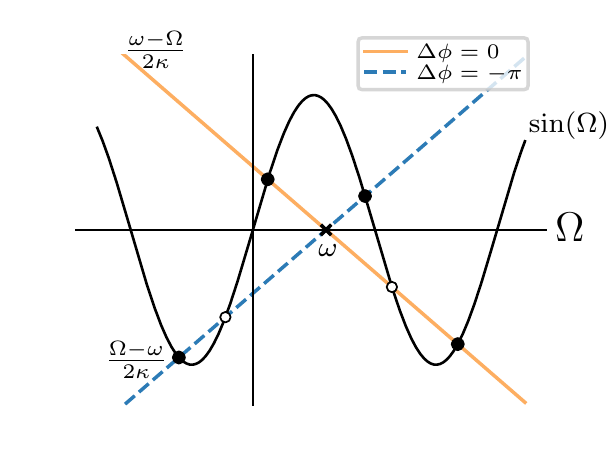}
	\caption{\label{fig.1}
	Graphical solution of Eq.~(\ref{eq.freq_rearrange}) for $m=\{0,-N/2\}$ corresponding to $\Delta\phi=\{0,-\pi\}$, respectively. Intersections of the sine and the linear functions represent the stable (filled dots) and unstable (hollow dots) solutions. The cross at $\Omega=\omega$ denotes the natural frequency of the oscillators.}
\end{figure}

In order to understand the stability criteria for these two states, it is convenient to solve Eq.~(\ref{eq.frequency}) graphically. Rearranging the equation yields:
\begin{equation}
\label{eq.freq_rearrange}
\frac{\omega-\Omega}{2\kappa\cos(\Delta\phi)}=\sin(\Omega)\,.
\end{equation}
In this form the solutions to this equation correspond to the intersection points of a linear and a sine function (see the sketch in Fig.~\ref{fig.1}). As proven in ~\cite{earl} the stability of the solutions is determined by the slope of the $\sin(\Omega)$ term which is exactly $\cos(\Omega)$.
%
%%%%%%%%%%%%%%%%%%%%%%%%%%%%%%%%%%%%%%%%%%%%%
%%%%%%%%%%%%%%%%%%%%%%%%%%%%%%%%%%%%%%%%%%%%%%
%

\begin{figure}
%\subfloat[$m=0$]{\includegraphics[width=0.5\linewidth]{plot0}} \subfloat[$m=-5$]{\includegraphics[width=0.5\linewidth]{plot5}}\\
%\subfloat[$m=1$]{\includegraphics[width=0.5\linewidth]{plot1}} \subfloat[$m=4$]{\includegraphics[width=0.5\linewidth]{plot4}}\\
%\subfloat[$m=2$]{\includegraphics[width=0.5\linewidth]{plot2}} \subfloat[$m=3$]{\includegraphics[width=0.5\linewidth]{plot3}}
\includegraphics[width=10cm]{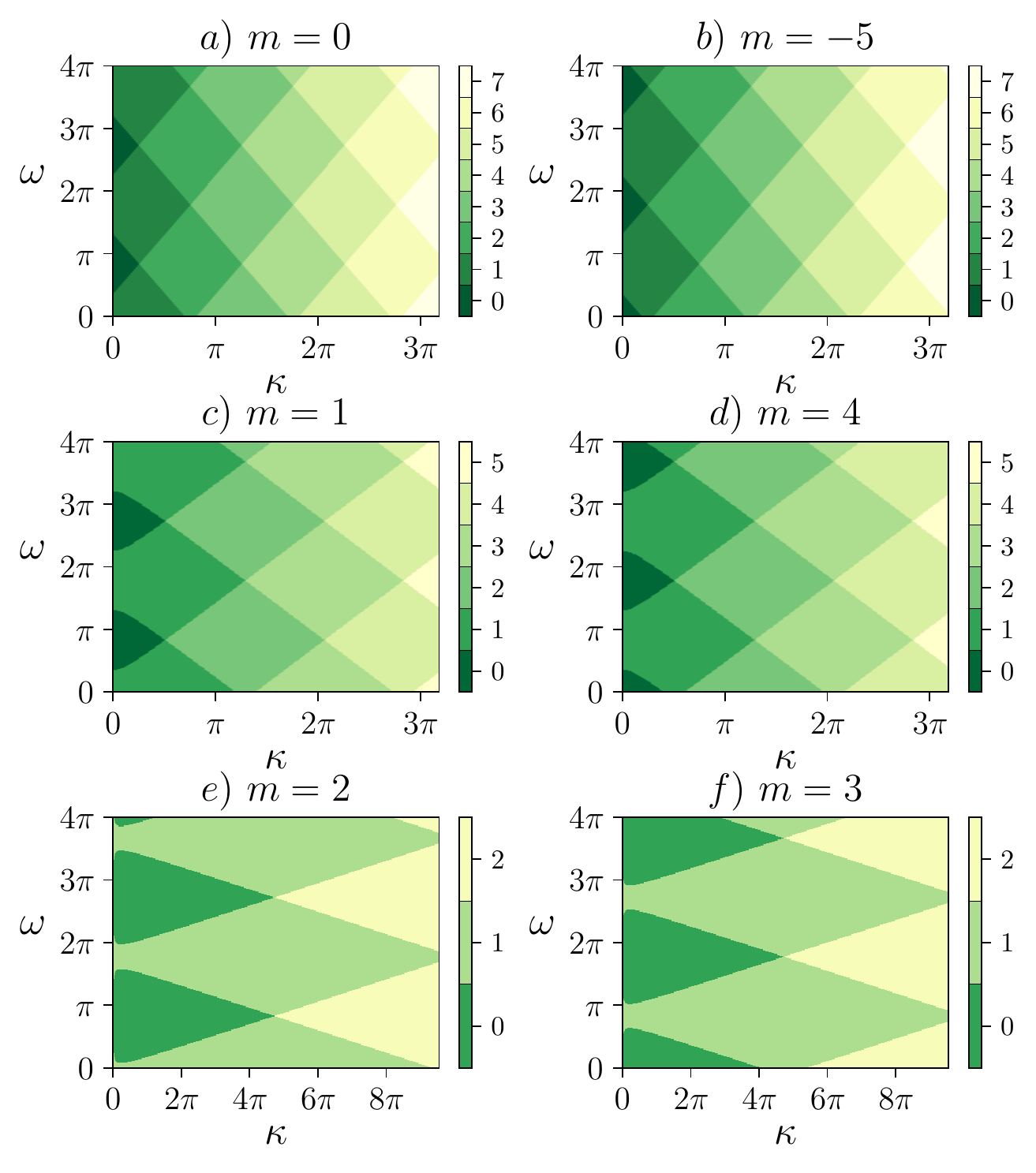}
\caption{Number of stable $\Omega$ solutions of Eq.~(\ref{eq.frequency}) for different states $m$ indicated by a colormap.  Stability is investigated as a function of $\omega$ and $\kappa$ parameters, by calculating numerically the exponents in Eq.~(\ref{eq.diagonal}).  Results for $m\in\{0,-5,1,4,2,3\}$ values are shown in panels $(a)-(f)$, respectively. The number of oscillators is $N=10$.}
\label{fig.2}
\end{figure}

In order to gain information about the other states we need to go deeper in the realm of Delay Differential Equations (DDE). The amplifying or diminishing trend for the considered $\eta_i(u)$ perturbations can be determined by a solution that is expressed with the matrix Lambert W function \cite{pease1965methods}. This analysis is done in the Appendix~\ref{appendix}. Using this method we can 
numerically investigate for any $m$ value the stability of the  $\Omega$ values obtained from Eq.~(\ref{eq.frequency}).   
Fig.~\ref{fig.2} shows the stability map of different states in the $\omega-\kappa$ space evaluated with the method from the Appendix~\ref{appendix} using the exponents in Eq.~(\ref{eq.diagonal}). 
More precisely, on the graphs from Fig.~\ref{fig.2} we indicate by a color-code the number of stable solutions with winding number $m$ for given $\omega$ and $\kappa$ parameters. Although the stability is determined by a complicated formula of transcendental functions, the stability map shows high level of symmetry and periodicity. First, each stability map is $2\pi$ periodic with respect to $\omega$. This is a result of the fact that an $\omega'=\omega+2\pi$ transformation will not change the form of Eq.~(\ref{eq.frequency}) and will increase the value of $\Omega$ with $2\pi$, leaving $\kappa_c$ and $\kappa_s$ invariant (notations introduced in Eqs.~(\ref{kappa_c}) and (\ref{eq.diagonal}) in the Appendix).  Second, the patterns on the stability maps are translated one relative to the other on the vertical axis by a value of $\pi$, for pairs where $|m_1|+|m_2|=N/2$ (states in the same row in Fig.~\ref{fig.2}). This  can be understood by the observation that the $\Omega'=\Omega-\pi$, 
$\omega'=\omega-\pi$, $\Delta\phi'=\Delta\phi+\pi$ transformations will leave Eq.~(\ref{eq.frequency}), and $\kappa_c$ and $\kappa_s$ in Eq.~(\ref{eq.diagonal}) invariant.

%%%%%%%%%%%%%%%%%%%%%%%%%%%%%%%%%%%%%%%%%%%%%
%%%%%%%%%%%%%%%%%%%%%%%%%%%%%%%%%%%%%%%%%%%%%
%

\section{Basins of attraction}
Linear stability analysis provides information about the dynamics of the system only in the proximity of the stationary states. For systems with dimensionality $d \geq 3$ the dynamics may become more complex, and one cannot rely only on these information when predicting the appearance of the phase-locked symmetric states~\cite{wiley_strogatz_2006}. In order to deal with this problem we determine the relative size of the basins of attractions through numerical experiments by measuring the probability of appearance of a given state $m$. 

In the numerical experiments for the initialization of DDEs from Eq.~(\ref{eq.of_motion}) one needs to provide a set of functions, determining the state history of the system for $t\in[-\tau,0)$~\cite{wernecke2019}. 
Here we consider an experimentally feasible procedure, by kick-starting the time-delayed Kuramoto system with a set of simple (non-delayed) uncoupled phase oscillators. This is possible by initializing the system with uniformly distributed random phases, $\theta_i(0)\in[0,2\pi)$, and solving Eq.~(\ref{eq.of_motion}) numerically with $K=0$ for the first $t\in[-\tau,0)$ period, or equivalently in the rescaled parameters $\kappa=0$ for $u\in[-1,0)$. This type of initialization of the history is equivalent to an interaction with a finite speed of propagation, meaning that there is no interaction between the oscillators up to a certain point. The final state is verified using the generalized order parameter $r_m$ defined in~\cite{denes2019}:
\begin{equation}
r_m =\left| \frac{1}{N}\sum_{j=1}^N e^{i[\theta_j(t)-(j-1)\frac{2m\pi}{N}]} \right|\,.
\label{order-param}
\end{equation}
It has been shown in Ref.~\cite{denes2019} that as the system approaches to a stable state characterized by the winding number $m=m^*$ the corresponding order parameter converges to 1 ($r_{m^*}\rightarrow 1$), while all the others decline to 0 ($r_{m\neq m^*}\rightarrow 0$).
Therefore, the long term behavior, which is the final phase-locked state for a randomly selected initial condition, may be identified by tracking the evolution of the order parameters. 
Selecting a high enough threshold one may identify the final state with a high confidence as the one corresponding to $r_m\rightarrow 1$.
We note here that the stability analysis presented in the Appendix~\ref{appendix} suggests that the characteristic exponent governing the stability (Eq.~(\ref{eq.diagonal})) may have also nonzero imaginary part indicating that the system can exhibit oscillatory dynamics in the vicinity of the stable states. A prediction method based thus only on the derivatives of the order parameter proposed also in~\cite{denes2019} is not appropriate. 

%
%%%%%%%%%%%%%%%%%%%%%%%%%%%%%%%%%%%%%%%%%%%%%
\begin{figure*}
\includegraphics[width=\linewidth]{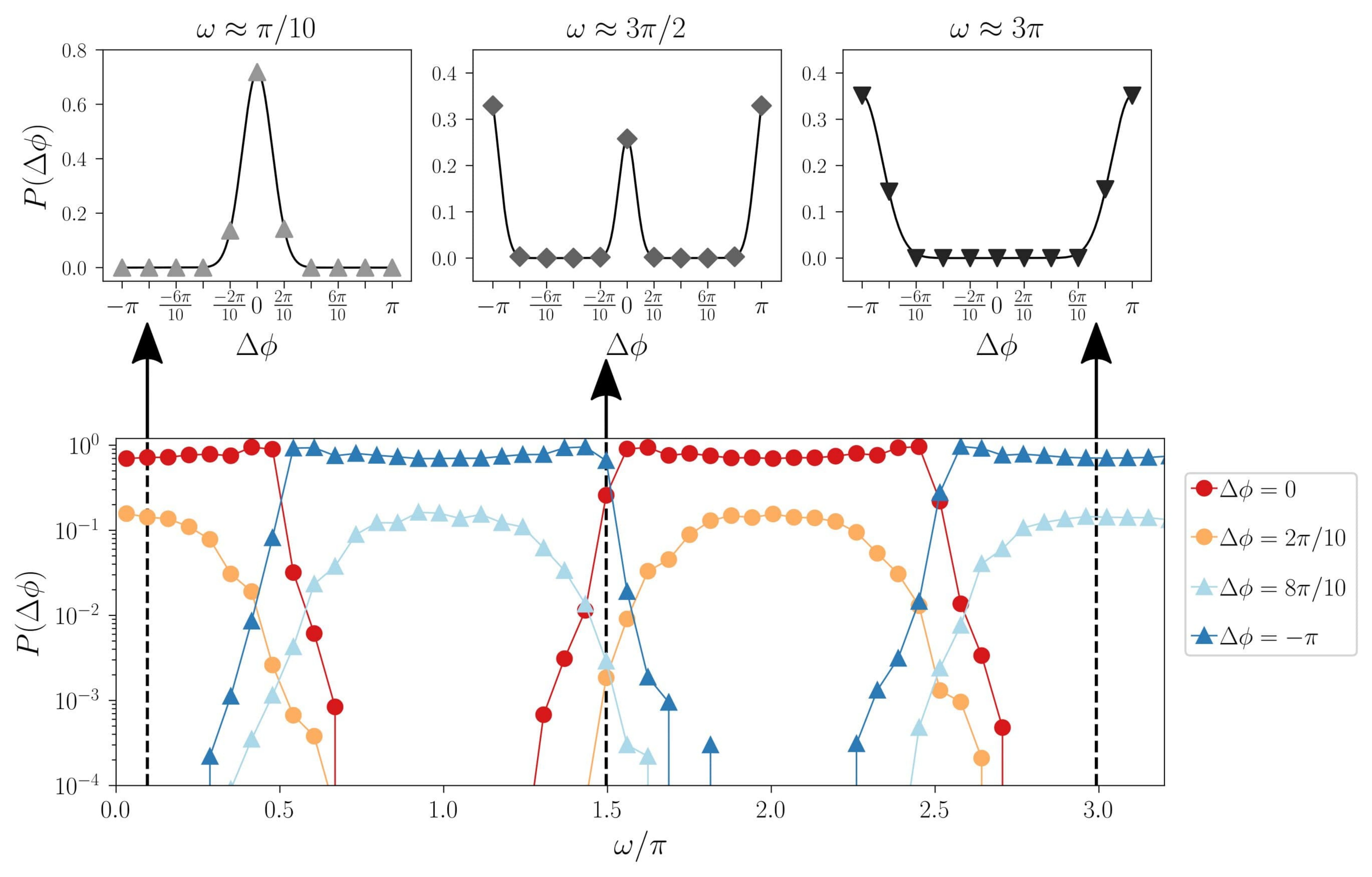}
\caption{\textit{Top:} Probability of the appearance of different phase-locked states for three $\omega$ values. \textit{Bottom:} Probability of the relevant phase locked states as a function of $\omega$. States with negative $m$ are omitted due to the symmetrical probability distribution function  with respect to 0. Dashed lines indicate the $\omega$ values corresponding to the probability distributions constructed on the top panel. The size of the system is $N=10$ oscillators, the coupling strength is $\kappa=2$.
\label{fig.3}}
\end{figure*}
%%%%%%%%%%%%%%%%%%%%%%%%%%%%%%%%%%%%%%%%%%%%%
%
Basin size distributions obtained from our sampling are illustrated in the top panel of Fig.~\ref{fig.3}. In order to interpret the graphs we remind that $m=0$ corresponds to $\Delta\phi=0$ phase difference and $|m|=5$ means $|\Delta\phi|=\pi$ anti-phase synchrony. A strong shift from the center to the sides can be observed in the distributions as $\omega$ changes. 
In order to avoid ambiguity, $\Delta\phi$ is defined as in~\cite{denes2019} such that there is no state with $\Delta\phi=\pi$. However, to get a symmetric distribution in Fig.~\ref{fig.3} we included it nevertheless by dividing equally the probability of the $\Delta\phi=-\pi$ state between $m=\pm5$, since at the level of the phase differences they correspond to the same state. The outcome of a more thorough analysis on the basin sizes is visible in the bottom plot of Fig.~\ref{fig.3}. As the results show, these distributions can be drastically different compared to the non-delayed case presented in~\cite{wiley_strogatz_2006,TILLES2011,denes2019}. The typical behavior of the system heavily depends on the parameters: for constant $\kappa$ there are intervals in $\omega$, roughly between $2k\pi-\frac{\pi}{2}$ and $2k\pi+\frac{\pi}{2}$, where the state $m=0$ is the most probable, while between $2(k+1)\pi-\frac{\pi}{2}$ and $2(k+1)\pi+\frac{\pi}{2}$ anti-phase synchrony is dominant. 

Taking into account that in case of the dimensionless equations the time delay is always one time unit, in order to understand the effect of the delay we need to switch back to the original parameter set. By keeping  $\omega_0$ and $K$ from Eq.~(\ref{eq.of_motion}) constant, one can conclude that this alternating behavior could be caused by the delay, making some states practically unobservable even though they are linearly stable (compare Fig.~\ref{fig.2} and Fig.~\ref{fig.3}). We note that this behavior is present if the system is initialized with random phases and the  history for initialization is considered as described previously. 
%
%%%%%%%%%%%%%%%%%%%%%%%%%%%%%%%%%%%%%%%%%%%%%
\begin{figure}
\includegraphics{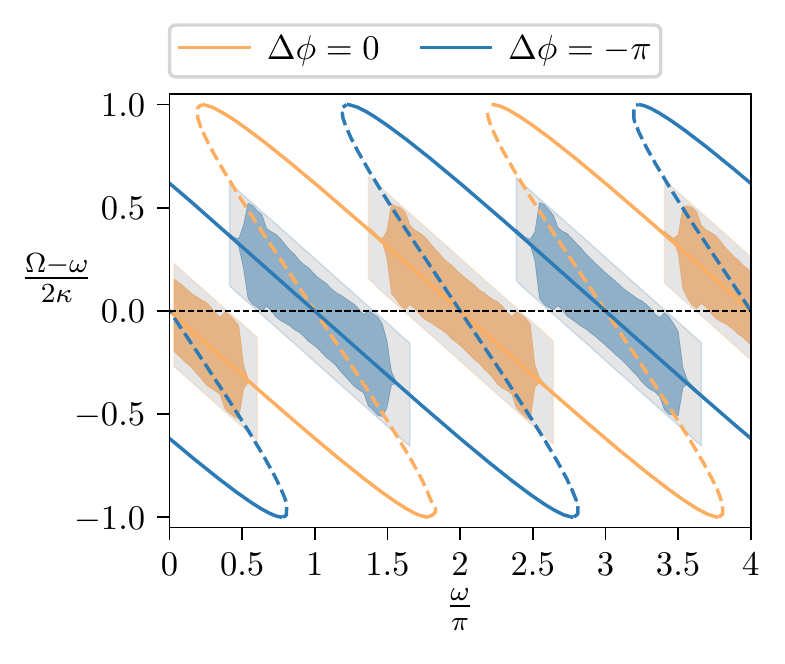}
\caption{Bifurcation diagram of the final frequencies for $m=0$ and $m=-5$ ($\Delta\phi=0$ and $\Delta\phi=-\pi$, respectively) as a function of the rescaled natural frequency $\omega/\pi$. On the vertical axis the difference between the final $\Omega$ and the natural $\omega$ frequencies is plotted, normalized with $2\kappa$. Stable/unstable frequencies are marked with solid/dashed lines. The black horizontal dashed line marks the $\Omega=\omega$ case. The width of orange and blue shadings correspond to the probability of observing the given frequency, compared to probability 1 indicated by gray shaded region. No shading at stable frequencies for any branch means zero observations. The other parameters are $N=10,~\kappa=2$.}
\label{fig.4}
\end{figure}
%%%%%%%%%%%%%%%%%%%%%%%%%%%%%%%%%%%%%%%%%%%%%
%
\section{Discussion}

It is interesting to correlate the changes in the $P(\Delta\Phi)$ probabilities to the bifurcations of the final frequencies as $\omega$ changes. As we learn from the $\omega-\kappa$ stability maps from Fig.~\ref{fig.2}, for most of the parameters the system is multistable, i.e.\ there are coexisting stable frequencies for a given parameter set. The effect of changing the parameters on the number and stability of $\Omega$ can be understood by solving Eq.~(\ref{eq.freq_rearrange}) graphically like in Fig.~\ref{fig.1}. Increasing $\kappa$ will make the slope of the linear function smaller in absolute value, thus there will be more intersections with the $\sin(\Omega)$ curve. For different $\Delta\phi$ values we will have lines with different slopes, all of them lying between the two extremes $\Delta \phi=0$ and $\Delta \phi =\pi$ and crossing each other at the $\Omega=\omega$ point. Modifying the $\omega$ is equivalent thus to sliding the linear function parallel to itself. Since the sine function is periodic, the number of solutions of Eq.~(\ref{eq.frequency}) changes with the same $2\pi$ periodicity. Even though the number of possible solutions is known, their stability, apart of $\Delta\phi=\{0,-\pi\}$,  is not yet determined.
For the full bifurcation diagram we need again Eq.~(\ref{eq.diagonal}) for sorting out unstable states. 

The bifurcation diagram of the two dominant states ($\Delta \phi=0$ and $\Delta \phi=-\pi$) is presented on Fig.~\ref{fig.4}. For comparison we also indicate the probability of these states by shaded areas. For the chosen $\kappa$ both states have at least one stable $\Omega$ frequency for all $\omega$ values, please compare this with Fig.~\ref{fig.2}. Though there exist multiple stable frequencies, only one of them appears almost exclusively as asymptotic state, given the initialization described above. Interestingly, as Fig.~\ref{fig.4} shows, the probability of the states strongly correlates with the difference between the dimensionless natural frequency $\omega$ and the final frequency $\Omega$. In other words, the system prefers a stable $\{\Delta \phi=0,\Omega_0\}$ state over an also stable $\{\Delta \phi=-\pi,\Omega_{-\pi}\}$ when $|\Omega_0-\omega|<|\Omega_{-\pi}-\omega|$ and vice versa. Furthermore, this dominance is not only relative, but the other state does not appear at all, except for narrow $\omega$ intervals around $(2k+1)\frac{\pi}{2}$, where the distances to $\omega$ are almost the same for both of the states mentioned before. A similar correlation between the resident time spent in one state and the difference in frequency  is reported in Ref.~\cite{DHUYS2014} in the case of delay coupled rings of oscillators with Gaussian white noise. Namely, the system spends the most time on orbits which have the frequency closest to the natural frequency of the rotators.

%
%%%%%%%%%%%%%%%%%%%%%%%%%%%%%%%%%%%%%%%%%%%%%
\begin{figure*}
\includegraphics[width=\linewidth]{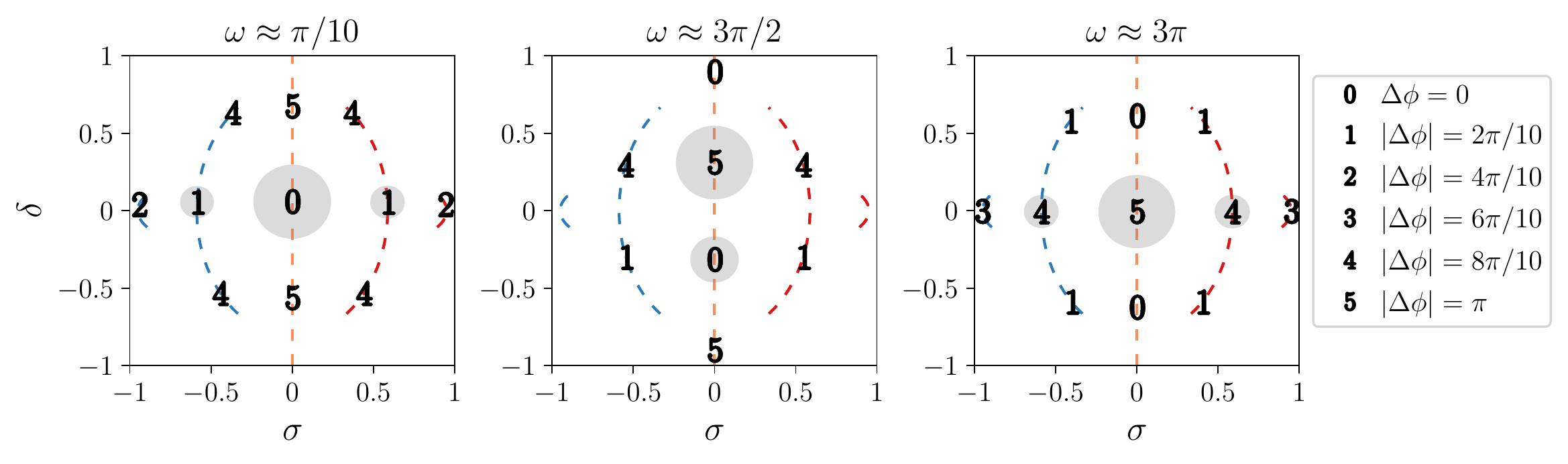}
\caption{Diagram of $\delta$ versus $\sigma$ for stable solutions of Eq.~(\ref{eq.frequency}) calculated for three values of $\omega$. Dashed lines and curves represent the paths along which the $\Omega$ frequencies for different states move as $\omega$ changes. Numbers, corresponding to $|m|$, indicate the $\sigma$ and $\delta$ values determined using actual final frequencies for the $\omega$ noted above. The area of the gray circles are proportional to the probability of a given $\{\Delta\phi,\Omega\}$ pair. Note that in the middle plot the states denoted by numbers 5 and 0 correspond to slightly different $\delta$ values, hence leading to different probabilities (compare Fig.~\ref{fig.3}). The parameters are $N=10,~\kappa=2.0$.}
\label{fig.5}
\end{figure*}
%%%%%%%%%%%%%%%%%%%%%%%%%%%%%%%%%%%%%%%%%%%%%
%
Let us consider now some heuristic arguments concerning the realization of other stable states different from $\Delta \phi=0$ and $\Delta \phi=-\pi$.
By recalling Eq.~(\ref{eq.frequency}) one can see that difference $\Omega-\omega$ is determined by
\begin{equation}
\label{eq.delta}
\delta=\cos(\Delta\phi)\sin(\Omega)\,,
\end{equation}
where the $-2\kappa$ factor is dropped to keep this parameter normalized. Calculating $\delta$ using the same $(\omega,~\kappa)$ parameters for other states $m$ than 0 and $-N/2$, one finds that $\delta$ can be closer to 0 than in the case of these two particular states. On the other hand, since the probability of these intermediate states tends to be smaller (see Fig.~\ref{fig.3}), there must be other relevant parameters as well in the process of state selection. One can notice that the main difference relative to the 
$\Delta \phi=\{0,-\pi\}$ case is related to the symmetry of the two sine terms in Eq.~(\ref{eq.rescaled}), taking different values if $\Delta\phi $ is neither 0, nor $-\pi$. One consequence of this is that in such states there is a preferred direction or order in which the oscillators arrange themselves, thus the symmetry of the interaction intensity for the two directions is broken. This is  mainly caused by the fact that sine is an odd function, so it is sensible to the interchange of oscillators and by numbering them we already imposed a predefined direction in which the phases should be compared to each other. In order to quantify the symmetry of a state, we introduce a second parameter
\begin{equation}
\label{eq.sigma}
\sigma=\sin(\Delta\phi)\cos(\Omega)\,,
\end{equation}
obtained by subtracting the two sine terms from Eq.~(\ref{eq.rescaled}) for a given $\Omega$ and $\Delta\phi$. Similarly to $\delta$ the $2\kappa$ term is omitted. Symmetric states will have $\sigma=0$.  As in case of the stability maps in Fig.~\ref{fig.2}, pairs of states with $|m_1|+|m_2|=N/2$ have similar trends of $\delta$ and $\sigma$ as $\omega$ changes. Also, pairs with opposite index ($m_1=-m_2$) have the same $\delta$, but opposite $\sigma$.  

In order to better understand the interplay between $\delta$ and $\sigma$ we plotted on Fig.~\ref{fig.5} the stable solutions of Eq.~(\ref{eq.frequency}) on the $\delta-\sigma$ plane for different $\omega$ values. Based on these graphs we can formulate some simple heuristic guidelines along which one can understand how the oscillators self-organize themselves.
\begin{itemize}
\item First, the selected state tends to be symmetric as possible, which explains why the peak of the distributions are always around 0 and $-\pi$. 
\item If there is more than one solution with the same $|\sigma|$ value, the one with the smaller $|\delta|$ will be preferred (see for example the $\sigma=0$ line). 
\item States, having the same $|\sigma|$ and $|\delta|$ are equally probable. 
\end{itemize}
As a result of the above observations, for a random ensemble of initial states and a given parameter set the calculation of the $|\sigma|$ and $|\delta|$ values could help us in estimating the expected attraction basin sizes. 

Finally, please note that our system is deterministic, so a probabilistic approach on the basin sizes should not be necessary, nevertheless the presence of time delay shifts the problem in the realm of infinite-dimensional systems~\cite{wernecke2019}, which limits the predictability of the dynamics. The $N$-dimensional phase space also looses its traditional meaning, since the $\theta_i$ values are not enough for a complete definition of a state in the phase space, the state histories are also needed.

\section{Conclusions \& outlook}
We studied symmetric phase locked states in rings of locally coupled Kuramoto rotators, interacting through time-delayed coupling. By symmetric phase-locked states 
we imply that a constant phase difference is allowed between neighboring oscillators. Former studies of such locally coupled systems determined a stability criterion for only a limited number of such states~\cite{earl,DHUYS2008}. An implicit stability condition is derived for all the states with constant phase difference. In order to understand the statistics for the selection of the stable stationary states, the attraction basins were investigated by considering large ensembles of uniformly distributed random initial states. Our numerical results show, that the size of the attraction basins can vary between several orders of magnitudes as a function of the parameters. This leads to the situation where stable stationary states are practically unobservable from random initial conditions due to their almost negligible attraction domain. By considering simple symmetry arguments we conjecture heuristic rules regarding the relative sizes of the attraction basins. During our numerical studies, in agreement with~\cite{wernecke2019}, we also observed that the presence of time-delay may often lead to the destabilization of simple fixpoint attractors, allowing the emergence of more complex dynamical behaviors besides these phase locked states. Searching  for different types of attractors in locally coupled time-delayed systems of oscillators should be thus another objective for future studies.

\begin{acknowledgments}
The present work was funded by the Romanian \mbox{UEFISCDI} grant nr. PN-III-P4-PCE-2016-0363.
\end{acknowledgments}

\appendix*
\section{Stability of linear DDEs using the Lambert W function\label{appendix} }
Substituting the proposed perturbed solution from Eq.~(\ref{eq.perturbation}) into Eq.~(\ref{eq.rescaled}) and keeping terms up to linear order yields the following linear DDE
\begin{equation}
\label{eq.dyn_perturbation}
\frac{\mathrm{d}\boldsymbol{\eta}(u)}{\mathrm{d}u}=\mathbf{A}\cdot\boldsymbol{\eta}(u)+\mathbf{A_d}\cdot\boldsymbol{\eta}(u-1),
\end{equation}
where $\boldsymbol{\eta}(u)$ is the perturbation vector, while $\mathbf{A}$ and $\mathbf{A_d}$ are $N\times N$ coefficient matrices. Matrix $\mathbf{A}$ is proportional to the identity matrix
 \begin{equation}
\label{eq.matrix_A}
\mathbf{A}=-\kappa\left(\cos(\Delta\phi+\Omega)+\cos(\Delta\phi-\Omega)\right)\cdot\mathbf{1}_N\,,
\end{equation}
while matrix $\mathbf{A_d}$ is a circulant matrix of the form
\begin{widetext}
\begin{equation}
 \mathbf{A_d}= \kappa \cdot
 \begin{pmatrix}
0	&	\cos(\Delta\phi-\Omega)	& 0     &		\dots		&     0    &	   \cos(\Delta\phi+\Omega)\\
\cos(\Delta\phi+\Omega)	&	0	&	\cos(\Delta\phi-\Omega) &  	\dots  	&	 0   & 	  0\\
0   &   \cos(\Delta\phi+\Omega)	&	0	& 	\dots  	&	 0   & 	  0\\
\vdots	&	\vdots	&		\vdots	&		\ddots	&   \vdots   &  \vdots \\
0	&	0	&	0  &   	\dots	&	0     &	      \cos(\Delta\phi-\Omega)\\
\cos(\Delta\phi-\Omega)	&	0	&  0   &	\dots	&  \cos(\Delta\phi+\Omega)   & 0\\
 \end{pmatrix},
\label{eq.matrix_A_d}
\end{equation}
\end{widetext}
where $\mathbf{1}_N$ is the $N\times N$ identity matrix. Eq.~(\ref{eq.dyn_perturbation}) is a system of linear DDEs. For systems where the matrices of the delayed and non-delayed terms commute the solution is known~\cite{asl2003analysis}.  In our case the commutation of $\mathbf{A}$ and $\mathbf{A_d}$ is fulfilled since $\mathbf{A}$ is proportional to the identity matrix. The solution hence writes as:
\begin{equation}
\label{eq.solution}
\boldsymbol{\eta}(u)=\sum_{k=-\infty}^{+\infty}\mathrm{e}^{\left[\mathbf{W}_k(\mathbf{A_d}\mathrm{e}^{-\mathbf{A}})+\mathbf{A}\right ]u}\mathbf{c}_k^I\,.
\end{equation}
The matrix $\mathbf{W}_k$ is $k$-th branch of the matrix Lambert W function~\cite{pease1965methods}, while the vector $\mathbf{c}_k^I$ depends on the coefficient matrices, the initial conditions, and the history of the delayed system.
Generally, in order to determine the stability we have to calculate the greatest eigenvalue of the exponent in square brackets over all $k$ values. 
Since the Lambert W function has infinite branches the whole eigenspectrum of the linear solution is infinite. For a scalar system, as opposed to Eq.~(\ref{eq.rescaled}), the principal branch of $k=0$ gives the rightmost eigenvalue. Fortunately for us the above statement holds for systems of DDEs as well if the coefficient matrices commute~\cite{asl2003analysis}.
The term that contains the principal branch is the following:
\begin{equation}
\label{eq.branch_0}
\boldsymbol{\eta}_0(u)=\mathrm{e}^{\left[\mathbf{W}_0(\mathbf{A_d}\mathrm{e}^{-\mathbf{A}})+\mathbf{A}\right ]u}\mathbf{c}_0^I\,.
\end{equation}
In order to easily calculate matrix exponentials we need to diagonalize the exponent. First, we have a matrix exponential in the argument of the Lambert W function, which is easy to calculate since $\mathbf{A}$ is diagonal:
\begin{equation}
\label{eq.exp_A}
\mathrm{e}^{-\mathbf{A}}=\mathrm{e}^{2\kappa\cos(\Delta\phi)\cos(\Omega)}\cdot\mathbf{1}_N.
\end{equation}
Thus the argument of the matrix Lambert W function is simplified, having only one matrix multiplied by a scalar
\begin{equation}
\label{kappa_c}
\mathbf{W}_0(\mathbf{A_d}\mathrm{e}^{-\mathbf{A}})=\mathbf{W}_0(\mathrm{e}^{2\kappa_c}\mathbf{A_d}),
\end{equation}
where we introduced the shorthand notation $\kappa_c=\kappa\cos(\Delta\phi)\cos(\Omega)$. The $\mathbf{A_d}$ matrix is circulant, thus diagonalizable using the unitary discrete Fourier transform  matrix~\cite{davis1979circulant}: $\mathbf{A_d}=\mathbf{U}\cdot\mathbf{D}\cdot\mathbf{U}^{-1}$. This property helps in evaluating the matrix Lambert W function, which is simply the scalar function acting on the diagonal elements
\begin{eqnarray}
&\mathbf{W}_0(\mathrm{e}^{2\kappa_c}\mathbf{A_d})=\nonumber\\&\mathbf{U}\cdot
\begin{pmatrix}
W_0(\mathrm{e}^{2\kappa_c}\lambda^1) & 0    &0 \\
0 & W_0(\mathrm{e}^{2\kappa_c}\lambda^2)    &0 \\
\vdots & \vdots   &\vdots \\
0 & 0   & W_0(\mathrm{e}^{2\kappa_c}\lambda^N)
\end{pmatrix}
\cdot\mathbf{U}^{-1},\nonumber\\
\label{eq.matrix_lambert}
\end{eqnarray}
where $\lambda^i$ is the $i$th eigenvalue of $\mathbf{A_d}$. For compactness we denote the diagonal matrix from above as $\mathbf{D}_W(\mathrm{e}^{2\kappa_c}\lambda)$. In Eq.~(\ref{eq.branch_0}) we have a sum in the exponent which can be decomposed to product of exponentials, since $\mathbf{A}$ commutes with the triple product in Eq.~(\ref{eq.matrix_lambert}):
\begin{equation}
\label{eq.sum_exponential}
\mathrm{e}^{\left (\mathbf{U}\cdot\mathbf{D}_W(\mathrm{e}^{2\kappa_c}\lambda)\cdot\mathbf{U}^{-1}+\mathbf{A}\right)u}=\mathrm{e}^{\mathbf{U}\cdot\mathbf{D}_W(\mathrm{e}^{2\kappa_c}\lambda)\cdot\mathbf{U}^{-1}u}\cdot\mathrm{e}^{\mathbf{A}u}.
\end{equation}
We can easily rearrange the terms by evaluating the first exponential using expansion in power series:
\begin{equation}
\label{eq.exp_taylor}
\mathrm{e}^{\mathbf{U}\cdot \mathbf{D}_W\cdot\mathbf{U}^{-1}u}
=\mathbf{U} \sum_{k=0}^{\infty}\frac{1}{k!}\left(\mathbf{D}_Wu\right)^{k}\mathbf{U}^{-1}%
=\mathbf{U}\mathrm{e}^{\mathbf{D}_Wu}\mathbf{U}^{-1}.
\end{equation}
Since $\mathbf{U}^{-1}$ commutes with the second exponential in Eq.~(\ref{eq.sum_exponential}) we arrive to the following:
\begin{equation}
\mathrm{e}^{\left (\mathbf{U}\cdot\mathbf{D}_W(\mathrm{e}^{2\kappa_c}\lambda)\cdot\mathbf{U}^{-1}+\mathbf{A}\right)u}=
\mathbf{U}\cdot\mathrm{e}^{\mathbf{D}_W(\mathrm{e}^{2\kappa_c}\lambda)u}\cdot\mathrm{e}^{-2\kappa_c\mathbf{1}_Nu}\cdot \mathbf{U}^{-1}.
\end{equation}
The exponents of the exponential functions commute again and their sum is a diagonal matrix, thus we can finally rewrite Eq.~(\ref{eq.branch_0}) as follows
\begin{equation}
\label{eq.sol_branc_0}
\boldsymbol{\eta}_0(u)=\mathbf{U}\cdot \mathbf{M}(u) \cdot\mathbf{U}^{-1}\mathbf{c}_0^I,
\end{equation}
where $\mathbf{M}(u)$ is a diagonal matrix with entries
\begin{equation}
\mathbf{M}_{jj}(u)=\mathrm{e}^{\left[   W_0\left( 2(\kappa_c\cos\frac{2\pi j}{N} + \mathrm{i}\kappa_s\sin\frac{2\pi j}{N} ) \mathrm{e}^{2\kappa_c} \right)   -2\kappa_c \right]u},
\label{eq.diagonal}
\end{equation}
where $j=0,1,...~N-1.$
Here we already inserted the eigenvalues of $\mathbf{A_d}$ and used the notation $\kappa_s=\kappa\sin(\Delta\phi)\sin(\Omega)$ similar to $\kappa_c$. The dynamics of the $\eta_i$ perturbations is governed by the linear combination of such terms, meaning that if the real part of all $N$ exponents is negative then the phase locked state for a given $\{\Omega,\Delta\phi\}$ pair is linearly stable.  

\bibliography{Kuramoto_delay.bib} 
\bibliographystyle{abbrv}

%\bibliography{Kuramoto_delay}
\end{document}